# Risk Management of Unmanned Aerial Vehicles

Hamid Reza Naji
*Dept. of Computer Engineering and Information Technology,
Graduate University of Advanced Technology,*
Kerman, Iran
naji@kgut.ac.ir

Aref Ayati
*Dept. of Computer Engineering and Information Technology,
Graduate University of Advanced Technology,*
Kerman, Iran
a.ayati@kgut.ac.ir

*Abstract*— **This paper presents an efficient risk management model for unmanned aerial vehicles or UAVs. Our proposed risk management establishes a cyclic model with a continuous and iterative structure that is very adaptable to agile methods and all IT-related resources. This model can be used in many applications, but as a case study, we have discussed it for UAVs. The increasing use of UAVs or drones in many fields and the existence of different threats is the main reason to have an efficient risk management method for them. In this paper, we cover risks based on IT-driven assets to decrease the chance of losing any data, failing the equipment or the system, and missing the reputation or credit based on cyclic and iterative flow. Our current offered risk management model for UAVs or drones is based on qualitative measures and can cover most of IT-based risks.**

## I. INTRODUCTION

Autonomous or remote-control vehicles called drones are increasingly used every day. Remote control vehicles can be Unmanned Aerial Vehicles (UAVs)), boats, ships and etc. [1]. Based on research by USA's FAA (Federal Aviation Administration), in these years, drones technology is becoming popular because the number of drones consumers increased from 1.1 million to 3.55 2021 in compare to 2016 [2], and also the US military had a significant investment (from 2.3 to 4.2 billion) in this field between 2008 to 2012 for research and production [3]. Referring to these researches, we can mention that the use of drones for different applications in organizations is rapidly increasing.

Risk management is also one of the main steps nearly in all industrial projects and all organizations' processes. In the drone field, based on their importance and their affiliated organization, we need to implement a complete risk management method. Risk Management should be considered as an ongoing process of assessing the likelihood of risks that are exists in any organization and measures to remove or to reduce the risks should be planed [4].

In this paper, we look at the risks of drones as a case study and base on that implement our risk management framework. Suppose that a postal UAV has a Ground Control Station called GCS with some resources such as Cloud (or Servers), and a network for communication between them. These "assets or resources" require risk identification in the first place. In a small scale, we know that a postal UAV has some hardware parts such as sensors, actuators, and some software parts such as codes, protocols, etc. We want to try identify them with a risky mindset and then try to assess and evaluate risks and get the best decisions with solutions to manage them. Additionally, we must know drone capabilities and limitations, such as how far they can go, in which weather they can go. How much the maximum distance between them and GCS should be and so on.

We have used some recently published surveys to get threats that we need for risk assessment. In the first step, we have used the [5] and [6] as these articles provided a list of the drone-associated and IoT threats in recent years. And in the second step, we used the [7] because this article attempted to document the cloud computing threats and issues. Also, we have used some other resources for providing the data and information we need for our research.

Note that we present a framework for risk management with a flow for managing those risks that any drone could expose, based on an idea of ISMS (Information Security Management System) developed by Enisa (European Union Agency for Cybersecurity) [8]. In fact, our offered framework can adapt to many areas including drones.

As [9] mentions, we tend to minimize the effect of any risk and maximize our benefits. Some risks are so insignificant that they can be ignored, but some should be considered. You have to consider that we are not looking for specific best practices in this area as we don't want to have one unique solution. In each risk case, we have many solutions that we need to decided based on our limitations, and we like to make a risky mindset in the affiliated organization [10].

## II. RELATED WORKS

In recent years, many papers tried to cover drone area risk management. For example, researchers in [11] introduced a safety assessment process model for UAVs with Petri nets, and in [12], researchers found human factor errors. They tried to present, identify and mitigate a systems for it. In [13], we see a risk assessment for small unmanned aircraft systems. in [14], the researcher talks about risks, vulnerabilities, and safety of UAVs in the transport system industry. In [15] we can find an excellent method for risk assessment in UAV operations. In [16] the researchers made an excellent qualitative and quantitative risk analysis with Bayesian Network (BN) based on ISO-12100 and ISO-13849 on UAVs. Still, in all of this researches, we didn't see any special try to build a quick risk management framework based on information Technology (IT) assets and Business Continuity Management (BCM). Here, we are trying to cover these issues in our paper.

### III. Prerequisites Of Our Proposed Framework

Before presenting the proposed framework, we will have some required descriptions.

*A. Basic Definitions*

**Risk:** An event that can prevent our operation or business from reaching its goals.

**Risk Management:** The ongoing process of assessing and documenting the likelihood of an unwanted event happening, and implementing measures for reducing the risk from the outcome of this destructive event. Note that in this paper, we like to offer a new framework that is continuous (not ongoing) and iterative. This means the risk team should try to manage risks with the proposed flow, for example, every 2-4 weeks iteratively.

**Business Continuity Management (BCM):** The ability of each organization to keep the ongoing activities and services after any critical event or incident.

Keep in mind that we want to simultaneously cover the BCM with IT risk management in this paper.

*B. Steps of Our Risk Management Framework for Drones:*

In our risk Framework, we present five steps with their dependencies. Keep in mind that our offered framework has a continuous and iterative flow. The critical thing in this framework is a documentation part at each step of the framework. It means that after any action, documentation is very important step because this can help other members of the team or organization understand the details.

**Making Risk Profile:** This is the first step of the framework that we want to present to the affiliated organization; the description of any risk and its type, and which IT assets were affected in the consequence of happened risks. This profile will be developed from cooperation between the business owner and the security risk team. In this step, the risk type associated with external (E) and internal (I) options is what we wish to describe. Any risk that originates from outside the organization by an external agent is considered to be external type from our perspective. However, the internal type originates from an insider at the company.

**Risk Assessment:** This is the second step in the framework that will want to discuss the risk with their granularity. The central part of this step is risk rating which comes from the risk matrix that is presented in continue. There are other components that are concerned with the threat that can create risk and its source, as well as the vulnerability that the threat source can utilize to create the danger, and also, we have a security impact part and the set of severity and likelihood of each risk. This part is the job for the security risk team. Note that in the part of security impact, we will find any risk impact on the C-I-A-A quadruple: confidentiality, Integrity, Availability, and Accountability [17].

**Confidentiality:** Guarantee that information is kept secret from unauthorized people, procedures, or equipment.

**Integrity:** Safeguard against unauthorized production, alteration, or elimination of information.

**Availability:** Prompt, dependable access to data and information services for authorized users.

**Accountability:** Process of tracking, or the capability to track, and actions to a liable source.

For example, if an opponent eavesdrops on any non-encrypted data transmitted between each asset, it's a risk to confidentiality. If an intruder wants to tamper with a message, it's a risk to integrity. If an intruder tries to jamming-attack on a UAV, he can make a risk on availability, Finally, if an intruder uses a method that is not traceable or trackable, it will impact accountability. In some cases, an intruder wants to try an attack that can impact on repudiations calls non-repudiations that they are also impacts on accountability. After the discussion about risks and their complexities, we need to rate each risk based on its severity and likelihood. For this work, two approaches are available which are Qualitative and Quantitative. in this paper, we'll focus on the qualitative risk assessment as it is simple to start with based on bellow options:

For Severity, we propose the four following measures that the assessor could map each risk based on the level of risk impact:

- **Critical (C):** This severity will present each risk that can impact the organization's foundation, especially reputation and monetary resources. Organizations must spend lots of money to improve these risks, but the essential tip is to lose some consumers forever.
- **High (H):** A risk that can cause failure in any operation or can cause failure in the organization's infrastructure or loss/deactivation of the organization's IT assets. The consequence of This type of risk is severe customer dissatisfaction, employees leaving the organization, and incorrect execution of every operation.
- **Moderate (M):** each risk categorized in M could cause any issues, problems, or disturbances in the correct and error-free execution of any operation that needs lots of time to solve. Results of this type can cause many customer complaints, asset destruction, or employee dissatisfaction.
- **Low (L):** Any risk placed in this category often does not cause much damage and problem. Most of these cases have small financial implications that can be ignored and do not seriously endanger the organization's goals.

For risk Likelihood, we propose four measures that the assessor could map each risk based on the probability of occurrence:

- **Negligible (N):** The likelihood of a risk is around one percent (1%); actually, in recent years it has not occurred.
- **Low (L):** The likelihood of a risk is around twenty-five percent (25%); actually, it has been seen less in recent years.
- **Moderate (M):** The likelihood of a risk is around fifty percent (50%); which means it can occur in recent months.
- **High (H):** The likelihood of a risk is around seventy-five percent (75%), means likely occurring these days.
- **Very-High (VH):** The likelihood of a risk is around a hundred percent (100%), and the chance of its occurrence is very high.

In the end, for rating and addressing the risks based on their importance, we should should use the risk matrix shown in TABLE I.

| Risk Ratings | | Severity | | | |
|---|---|---|---|---|---|
| | | C | H | M | L |
| Likelihood | VH | C | C | H | M |
| | H | C | C | H | L |
| | M | H | H | M | L |
| | L | M | M | L | L |
| | N | L | L | L | L |

TABLE I. Risk Rating Matrix

Depending on the conditions of the organization, the number of qualitative items in the probability scale and severity scale can be reduced or increased by the security risk team.

**Risk Evaluation:** After assessing risks, we must decide on the best solutions for risk exposure. In this step, the main discussion is about the decision and solution for each risk. In the decision part, we can make a policy based on four decision rules: *Accepting Risk, Risk Avoidance, Transferring the Risk,* and *Mitigating Risk*. In each case, the assessor can take one of them as a decision and get a solution based on coordinating with the business owner and other dependent persons to that risk.

In the **"Avoid** risk" decision, we want a plan to eliminate all risks.

In the "**Transfer** risk" decision, we want a plan to transfer a risk responsibility to others, such as insurance.

In the **"Mitigation** risk" decisions, we want a decreasing risk planning to minimize risk impact, such as setting up controls.

In the "**Accept** risk" decision, we have no choice but to accept or even control the risk.

**Risk Treatment:** In this step of the proposed framework, the critical thing is to make a mitigation plan and some controls for each risk and validate them. This should be done for each IT resource or asset in the organization and is charged to the custodian of each resource and security risk team.

**Monitoring and Audition:** The last step of the framework discusses monitoring and auditing. After implementing any treatment, we must monitor and audit it. This step should be controlled by the business owner and security risk team.

## IV. THE PROPOSED FRAMEWORK

After finishing the above steps, the first iteration of our flow is done. After the first step, the second step and then the BCM, should be implemented.

As shown in Fig.1 based on [18], we have shown our proposed framework with adequate granularity.

After assessing risks, the critical step is evaluation. In this step, the most important notice is to present logical, scientific, wise, and probable solutions. As a case, we want to see the postal organization that wants to send a package to the consumer's destination. In this example, if the transmitter of the drone has an issue on its Global Navigation Satellite System (GNSS) sending the drone's current location, what do you want to do? If Spoofing causes infecting your GPS, what do you do? do you want a high-quality GPS to install on the drone to eliminate the chance of this risk exposure? or do you have another offer? After thinking about this issue, maybe the better work is to have some alternative systems. in the original case, perhaps you have one GNSS, such as GPS. Still, you need alternatives such as Galileo, Beidou, and Glonass. In this status, deceiving the pilot or controller of the drone is becoming too complex. also, GPS Spoofing is much more complicated with good cryptography on the communication network between satellites and drones. This is the first solution that comes to mind. If we search for novel methods, we can find better ways.

One of the significant risks that you need to address is Human Resource Management (HRM) issues. Think that if you have a problem paying the developers' salary on time, they can make a backdoor on your drones, communications, or servers in the cloud system and leave the organization. After a while, they can use these backdoor codes made on your systems and try to attack your systems by the backdoor attack. This attack can impact the security quadruple of the C-I-A-A and infect the organization. Also, think about teaching employees. If you ignore this case, one of your employees can unintentionally infect by social engineering attack and install any malware on their device and harm the organization. By connecting devices to the organization's resources, the organization can also be infected. This case also can impact C-I-A-A. TABLE II. Risk Profile's List.

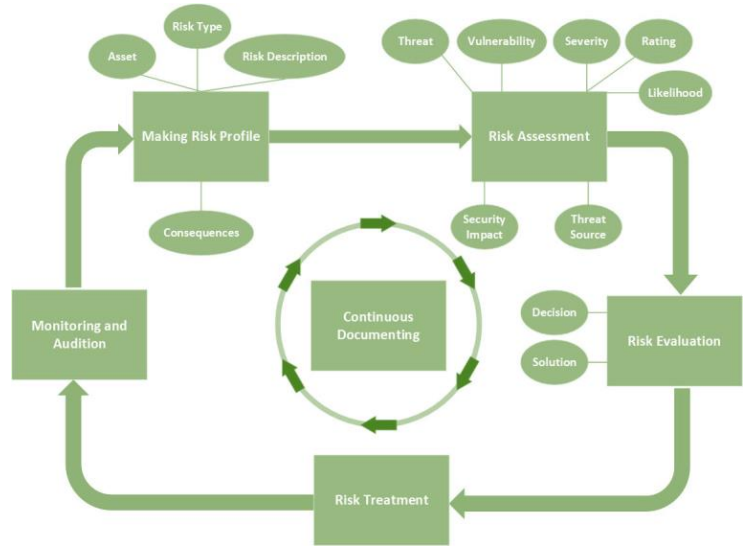

Fig. 1. Our Proposed IT Risk Framework for Drones is based on our similar previous paper that we are waiting for publish in the 12th Smart Grids Conference.

| | | **Risk Profile** | | | |
|---|---|---|---|---|---|
| Case # | Where(G/A) | Asset (Resource) | Risk Type | Risk Description | Consequence |
| 1 | A/G | All Nodes in the Network | I | Degrade Communication Quality | Network Problem |
| 2 | G | Cloud Resources | I | Becoming The Target of Any Elevation of Privilege on Access Control | Organization/ Operation |
| 3 | G | Client to Terminal network | E | Attempting to use weak Authentication mechanisms | Drone Loss/ Information leakage |
| 4 | A | Drone Nodes | I | Battery Packs Energy Getting Low or Destroy After a While | Limitation In Flight Duration |
| 5 | G | Human Resources | E | Becoming The Target of Social Engineering and Making Backdoors and Import Viruses into Systems | Organization/ Operation/ HR Damage |
| 6 | G | Operation | E/I | New and current Ethical/Legal Regulation Can Take Our Mission Impossible | Operation Problem |
| 7 | A/G | All Nodes in the Network | E | Attempting to Eavesdropping on transmitted information | Information leakage |

TABLE II. Risk Profile's List.

Any mistakes in security issues can result in a loss of the organization's reputation, and it can impact the BCM of the organization.

An important point that has not been paid much attention to in other risk management models and methods is that sometimes we face a serious challenge when facing a set of risks. Specifically, after deciding on them, some of them will have an equal value in the ranking. For example, suppose that we have five C-level risks, and the organization's security risk team will probably not be able to deal with all these five cases at the same time. Our suggestion in such cases is to use multivariate decision-making methods such as the Analytical Hierarchy Process (AHP) method [19]. In this case, for each criterion, the security team must introduce a number of sub-criteria for decision-making, and all the teams present their opinions based on the sub-criteria, and at the end, based on the weights given to these items. For lower-ranked cases, however, this method is not justified and will probably only take up the security team's time.

## V. A MINIMAL CASE STUDY

In this section, we want to present a minimal case study that can give you a better idea. Imagine that we have a postal organization that wants to send lots of packets or boxes to destinations with a drone. For IT risk management, we think one of the good ways to manage risks, especially for better documentation is to make a list is the best ideas. In our proposed list, we have three main columns with fourteen sub-columns. In the first main column, we offered to have five sub-columns for profiling risks. The first sub-column is known as "where," which discusses where a risk can happen. The second part is the "Asset" that can host a risk. In this sub-column, we have a series of options such as All nodes in the network, a part of the communication network, cloud resources, drones, human resources, and so on that can be an IT asset or resource. The third part is the "Risk Type" we discussed before. In the fourth part, we explain the risk description. And the last sub-column explains the risk outcome known as "Risk Consequence."

Now we have the base of risk profiling. Then we should go for risk assessment. In the risk assessment column, we offered seven sub-columns about assessing risks. The first one shows a vulnerability that a threat agent can use to pose a threat. The second one is "Threat," and the third one is "Threat Agent or Source." in this part of the list, we list the threats and agents that can make a risk for us. Security Impact (based on C-I-A-A ), Likelihood, Severity, and Rating are the last sub-columns of the Risk Assessment column discussed previously. For security impact, we showed impacts based on confidentiality with "C", Integrity with "I", Availability with "A", and Accountability with "a".

And finally, the last main column is "Risk Evaluation," and we discuss "Decisions" and "Solutions" in a separated sub-column of the risk evaluation. As we discussed previously, in the Decision sub-column, we like to see one of a foursome of Avoid/Mitigate/Transfer/Accept. And the last part of the list, we have "Solutions" based on logical, scientific, wise, and probable explanations.

Now we back to our example. We want to show you the offered list, part by part. As you can see in TABLE II. we have the main part of the risk profile with five sub-columns. In each row, we present one example of risk. We have six samples in this list. The first one is about the "degradation of communication quality" that can affect "all nodes in the network" with an internal agent and cause network problems. Other samples follow the same logic. However, it was mentioned earlier that we want to have an IT risk assessment with BCM, which is why we want to take some HR issues into the table.

In the second part, known as the risk assessment, we make a risk assessment list based on past risks. For example, based on the degraded quality of communication, we have a "lack of QoS" vulnerability, which outcome threats are packet loss/latency/ and so on. That is the outcome of miss knowledge or mistakes of developers. This risk can impact one of the C-I-A triads with high likelihood, Moderate severity, and high rate. Now you can see TABLE III. After TABLE III, we present TABLE IV, which shows the decision and solution sub-columns. Ultimately, we can combine all of these in a table.

| Risk Assessment | | | | | | | |
|---|---|---|---|---|---|---|---|
| Case # | Vulnerability | Threat | Source (Threat Agent) | Impact on | Likelihood | Severity | Rate |
| 1 | Lack of QoS | Packet Loss/Latency/Jitter/Congestion/Delays/Collisions | Development Team | I | H | M | H |
| 2 | Not Implementing the Right Policy for Access Control | Elevation/ Escalation of Privilege | Development/ Maintenance Team | CIAa | VH | M | H |
| 3 | Weakness in authentication Security | Password Cracking | Opponent | CIAa | H | C | C |
| 4 | Battery Packs | Battery Failure/ Expired Battery/ Charge Cycle Issues | maintenance team | A | L | C | M |
| 5 | Human Resource Challenges/ Employee Egos and dissatisfaction | Not Paying Attention to the salaries and benefits of employees | Malicious Insiders | CIA | L | M | L |
| 6 | Policy Changing and failure to comply with law | Law and Justice | Pilot/ Government | A | N | M | L |
| 7 | Weakness in Confidentiality Security | Man in the Middle or Eavesdropping Attacks | Opponent | C | H | H | C |

TABLE III. Risk Assessment's List.

For example, for the "Attempting to misuse weak authentication" risk that came from the "Password Cracking", we can use the solution that came in [20] that uses a Kerberos System for authentication. This will be an avoidance decision type and it is a logical, scientific, wise, and probable solution.

| Risk Evaluation | | |
|---|---|---|
| Case # | Decision | Solution |
| 1 | Mitigate | Increase the quality of the Development team and use QoS. |
| 2 | Avoid | Role-based access control (RBAC) can be implemented. |
| 3 | Avoid | Implementing Kerberos System [20]. |
| 4 | Mitigate | Implementation of the periodic check policy of parts. |
| 5 | Mitigate | Making a system for checking Malware infection based on Data science such as [21]. |
| 6 | Acceptance | This risk must be adopted. |
| 7 | Avoid | Using a Secure and Fast Cryptographic function. |

TABLE IV. Risk Evaluation List.

## VI. ANALYSIS WITH SWOT MATRIX

SWOT analysis is a technique of using a 2x2 square matrix that is used for management and strategic planning in an organization and is used to help a person or organization identify strengths, weaknesses, opportunities, and threats related to business competition or project planning. In the model of SWOT that we will use, we will seek to review and analyze our proposed framework for risk management. But by definition, SWOT examines the environment and competitiveness of an organization [22]. In TABLE V, a brief analysis for the proposed framework is provided which is a sample implementation in an active organization.

| |
|---|
| **Strengths:** <br> 1. Simple understanding of the framework and high level of maturity in documenting and understanding issues. <br> 2. It can clearly increase the speed of processes and security. <br> 3. Increases relationships and satisfaction between stakeholders. <br> 4. Scalability for use in different organizations. <br> 5. There is almost no flexibility in the framework. <br> 6. Compatibility with agile management frameworks. <br> 7. Guaranteeing the effectiveness of the measures taken. <br> 8. Quick coordination with the regulations that are set. <br> 9. Not recommending to use best practices. <br> 10.considered proposed framework with C-I-A-A tetrad. |
| **Weakness:** <br> 1. Low level of maturity in the work process. <br> 2. Weakness in managing people and communication due to the lack of specific roles and meaningful work procedures. <br> 3. Absence of a tool specific to this method for risk management. <br> 4. Higher cost due to continuous review and increased working hours. |
| **Threats:** <br> 1. Limited experiences in this field. <br> 2. Inability to create 100% security. |
| 3. Difficulty in understanding it 100% and lack of experience implementing it in organizations. |
| **Opportunities:**<br>1. Ease of expansion and integration with other frameworks.<br>2. Accepting the role of the security team in development teams instead of outsourcing security-related matters.<br>3. Simplicity due to similarity with ISO 27001, and NIST Cyber Security Framework. |

TABLE V. SWOT Matrix.

## VII. CONCLUSION AND FUTURE WORKS

This paper presented a risk management model for UAVs. Our proposed risk management model, while establishing a cyclical flow with a continuous and iterative structure, is very adaptable to agile methods and covers all IT-related resources in the organization.

Our proposed model, while supporting CUIs (Controlled Unclassified Information), is not based on the best practice and avoids them, forcing the security team members to study and analyze new and more effective defense methods. In fact, the security team should choose the most suitable new, wise and feasible methods to deal with any risk according to the conditions of their environment, and our proposal is such that at the same time as choosing a suitable method, high compliance with the rules should not be forgotten. As we considered in the paper, according to the results of using this risk management model in an IT-oriented system, the output of the development team in terms of the proper use of this model was very good, and the coordination of the team with this risk management method was high, as we observed in the presented SWOT. Also, we have suggested to use the AHP method for a more detailed ranking for those risks that are ranked the same.

## REFERENCES

[1] S. Herrick. "What's The Difference Between A Drone UAV and UAS?" Nov. https://botlink.com/blog/whats-the-difference-between-a-drone-uav-and-uas

[2] W. Atkinson. "Drones Are Gaining Popularity." ecmag. https://www.ecmag.com/magazine/articles/article-detail/your-business-drones-are-gaining-popularity

[3] L. C. Baldor, "Flashy drone strikes raise status of remote pilots," *The Boston Globe,* 2012.

[4] J. Slay and A. Koronios, *Information technology security and risk management*. Wiley, 2006.

[5] K.-Y. Tsao, T. Girdler, and V. G. Vassilakis, "A survey of cyber security threats and solutions for UAV communications and flying ad-hoc networks," *Ad Hoc Networks,* vol. 133, p. 102894, 2022.

[6] W. H. Hassan, "Current research on Internet of Things (IoT) security: A survey," *Computer networks,* vol. 148, pp. 283-294, 2019.

[7] H. Tabrizchi and M. Kuchaki Rafsanjani, "A survey on security challenges in cloud computing: issues, threats, and solutions," *The journal of supercomputing,* vol. 76, no. 12, pp. 9493-9532, 2020.

[8] Enisa. "The ISMS Framework." https://www.enisa.europa.eu/topics/threat-risk-management/risk-management/current-risk/risk-management-inventory/rm-isms/framework (accessed.

[9] A. Osborne, *Risk management made easy*. Bookboon, 2012.

[10] E. Wheeler, *Security risk management: Building an information security risk management program from the Ground Up*. Elsevier, 2011.

[11] P. Gonçalves, J. Sobral, and L. A. Ferreira, "Unmanned aerial vehicle safety assessment modelling through petri Nets," *Reliability Engineering & System Safety,* vol. 167, pp. 383-393, 2017.

[12] P. Neff and K. E. Garman, "Identifying and mitigating human factors errors in unmanned aircraft systems," in *16th AIAA Aviation Technology, Integration, and Operations Conference*, 2016, p. 3593.

[13] L. C. Barr *et al.*, "Preliminary risk assessment for small unmanned aircraft systems," in *17th AIAA Aviation Technology, Integration, and Operations Conference*, 2017, p. 3272.

[14] S. O. Johnsen and T. E. Evjemo, "State of the art of unmanned aircraft transport systems in industry related to risks, vulnerabilities and improvement of safety," in *Proceedings of the 29th European Safety and Reliability Conference (ESREL). 22–26 September 2019 Hannover, Germany*, 2019: Esrel 2019.

[15] K. Wackwitz and H. Boedecker, "Safety risk assessment for uav operation," *Drone Industry Insights, Safe Airspace Integration Project, Part One, Hamburg, Germany,* pp. 31-53, 2015.

[16] A. Allouch, A. Koubaa, M. Khalgui, and T. Abbes, "Qualitative and quantitative risk analysis and safety assessment of unmanned aerial vehicles missions over the internet," *IEEE Access,* vol. 7, pp. 53392-53410, 2019.

[17] M. A. Rahman, M. S. Abuludin, L. X. Yuan, M. S. Islam, and A. T. Asyhari, "EduChain: CIA-compliant blockchain for intelligent cyber defense of microservices in education industry 4.0," *IEEE Transactions on Industrial Informatics,* vol. 18, no. 3, pp. 1930-1938, 2021.

[18] S. A. Ayati and H. R. Naji, "A Novel IT-Based Lightweight Risk Management Framework for Metering Networks in Smart Grids," in *2022 12th Smart Grid Conference (SGC)*, 2022: IEEE, pp. 1-5.

[19] J. E. Leal, "AHP-express: A simplified version of the analytical hierarchy process method," *MethodsX,* vol. 7, p. 100748, 2020.

[20] S. A. Ayati and H. R. Naji, "A Secure mechanism to protect UAV communications," in *2022 9th Iranian Joint Congress on Fuzzy and Intelligent Systems (CFIS)*, 2022: IEEE, pp. 1-6.

[21] M. K. Alzaylaee, S. Y. Yerima, and S. Sezer, "DL-Droid: Deep learning based android malware detection using real devices," *Computers & Security,* vol. 89, p. 101663, 2020.

[22] E. GURL, "SWOT analysis: a theoretical review," 2017.

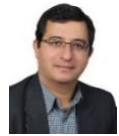

**Hamid Reza Naji** is associate professor of computer engineering at the Graduate University of Advanced Technology, Iran. His research interests include embedded systems, distributed, parallel and multi-agent systems, networks, and security. Dr. Naji has a Ph.D. in computer engineering from the University of Alabama in Huntsville, USA.

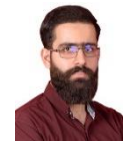

**Aref Ayati** graduated from Isfahan University with a B.Sc in computer sciences in 2020, and he earned an M.Sc degree in Information Technology Engineering in 2023 from Iran's Graduate University of Advanced Technology. His areas of interest include IT management, UAV Systems, and Computer Networks and security.